\documentclass[reprint, showpacs, superscriptaddress,nofootinbib]{revtex4-1}
\usepackage[utf8]{inputenc}
\usepackage[english]{babel}
\usepackage{amsmath}
\usepackage{amsfonts}
\usepackage{amssymb}
\usepackage{graphicx}
\usepackage{dcolumn}
\usepackage{booktabs}
\usepackage{afterpage}

\AtBeginDocument{
\heavyrulewidth=.08em
\lightrulewidth=.05em
\cmidrulewidth=.03em
\belowrulesep=.65ex
\belowbottomsep=0pt
\aboverulesep=.4ex
\abovetopsep=0pt
\cmidrulesep=\doublerulesep
\cmidrulekern=.5em
\defaultaddspace=.5em
} 
\begin{document}
\title{Theoretical study of electron structure of superheavy elements with an open $6d$ shell, Sg, Bh, Hs and Mt}
\author{B.G.C. Lackenby}
\affiliation{School of Physics, University of New South Wales,  Sydney 2052,  Australia}
\author{V.A. Dzuba}
\affiliation{School of Physics, University of New South Wales,  Sydney 2052,  Australia}
\author{V.V. Flambaum}
\affiliation{School of Physics, University of New South Wales,  Sydney 2052,  Australia}
\affiliation{Johannes Gutenberg-Universit\"at Mainz, 55099 Mainz, Germany}

\begin{abstract}
We use recently developed efficient versions of the configuration interaction method to perform {\em ab initio} calculations of the spectra
of superheavy elements seaborgium (Sg, $Z=106$),  bohrium (Bh, $Z=107$),  hassium (Hs, $Z=108$) and meitnerium (Mt, $Z=109$).
We calculate energy levels, ionization potentials, isotope shifts and electric dipole transition amplitudes.
Comparison with lighter analogs reveals significant differences caused by strong relativistic effects in superheavy elements.
Very large spin-orbit interaction distinguishes subshells containing orbitals with a definite total electron  angular momentum $j$.
This effect replaces Hund's rule holding for lighter elements.
\end{abstract}


\maketitle

\section{Introduction}

Theoretical study of electron structure of superheavy elements (SHE, nuclear charge $Z>103$) is an important area of research closing the gaps in relevant experimental data. While all SHE up to oganesson (Og, $Z=118$) have been synthesized and named \cite{Karol2016, HHO2013, OUL2004}, experimental data on their spectra are absent.

The heaviest elements for which experimental spectroscopic data are available are nobelium (No, $Z=102$)~\cite{Laatiaoui2016, Chhetri2018} and lawrencium (Lr, $Z=103$)~\cite{SAB15}. Ionization potential (IP) has been measured for both atoms and the frequency of the strong $7s^2 \ ^1$S$_0$ $\rightarrow$ $7s7p \ ^1$P$_1^{\rm o}$ electric dipole transition is measured for No. For heavier SHE the data come only from theory. There are many accurate calculations for atoms with relatively simple electronic structure, which includes atoms and ions with few electrons above closed shells (usually not more that four,
see, e.g. Refs.~\cite{eliav2015,pershina2015,E-E113,P-E113,E114a,E114b,E114c,E114d,E115,FF113-115,DD113-114}). This constitutes less then a half of the SHE in the range $104 \leq Z \leq 118$.
Most of the SHE have an open $6d$ or $7p$ shell with more than four electrons. Till recently, the only available tool to perform calculations for such systems was the multi-configuration Dirac-Fock method (MCDF, see, e.g. review~\cite{Fischer2016}). Some of the  MCDF results we discuss in section \ref{sec:SHEIP}. There are some model calculations of the basic parameters of the atoms, such as IP~\cite{Borsch115-117} and polarizabilities~\cite{Dzuba2016}. Accurate {\em ab initio} calculations of the spectra are practically absent. This is an unfortunate situation since from the study of relatively simple SHE we know that strong relativistic effects often bring significant difference in properties of SHE compared to their lighter analogs. Similar effects are expected for all SHE, including those with open shells. To address the problem,  we have developed efficient versions of the configuration interaction (CI) approach, which allows study of atoms with any number of valence electrons. This includes the so-called CIPT method (configuration interaction with perturbation theory,~\cite{DBHF2017}) and its fast version, the FCI method (fast configuration interaction~\cite{FCI}). Both methods are based on the idea that off-diagonal matrix elements between highly excited states can be neglected in the CI matrix. This allows one to reduce the problem to a much smaller matrix with modified matrix elements. The methods were tested on such open-shell systems as Yb and No (including states with excitations from the $4f$ and $5f$ subshells)~\cite{DBHF2017,FCI}, Ta~\cite{LDFDb2018}, W and I~\cite{DBHF2017}, and superheavy elements Db~\cite{LDFDb2018} and Og~\cite{LDF118}. 
Db is the first SHE with an open $6d$ shell that has been studied using the CIPT method~\cite{DBHF2017,Berengut-Db}. Its ground-state configuration is [Rn]$5f^{14}7s^25d^3$, i.e. it has five valence electrons above closed shells, which makes it difficult to use other methods. Lighter neighbours of Db, Rf (four valence electrons) and Lr (three valence electrons) were studied with the use of the powerful CI+all-order method~\cite{NoLrRf,IPNoLrRf}.
The use of the latter approach for Db is very problematic and is practically impossible for heavier elements. Following the successful use of the CIPT for Db, we apply it in the present work to heavier elements Sg, Bh, Hs and Mt ($106 \leq Z \leq 109$).

In this work we present the low-lying odd and even states of SHE $Z=106$-$109$ including the allowed E1 transition amplitudes and rates from the ground state to odd parity states. We also calculate the ionization potential and isotope shift parameters for these elements.\\

The paper progresses as follows; in Section \ref{sec:CIPT} we give a brief overview of the CIPT technique and how we implement it for the SHE. In Section \ref{sec:Accuracy} we discuss the accuracy of the calculations. In Section \ref{sec:Isoshift} we give a brief discussion on the calculation of E1 transitions and corresponding isotope shift parameters between synthesized and predicted meta-stable SHE. In sections  \ref{sec:Sg}, \ref{sec:Bh}, \ref{sec:Hs} and \ref{sec:Mt} we discuss the results of the CIPT on  Sg~\textsc{i}, Bh~\textsc{i}, Hs~\textsc{i} and Mt~\textsc{i} atoms respectively.  For reference we present the low-lying spectrum for Sg~\textsc{i} and Bh~\textsc{i} in Table \ref{tab:SHESpectrumSgBh} and Hs~\textsc{i} and Mt~\textsc{i} in Table \ref{tab:SHESpectrumHsMt}  and the E1 transitions and isotope shift parameters in  Table \ref{tab:SHEE1transitionSgBh}. In Section \ref{sec:SHEIP} we present the ionization potentials of the four elements and compare them with other calculations. 

\section{CIPT Method} \label{sec:CIPT}

As mentioned above, an open $6d-$shell with more than three  valence electrons makes established many-body methods too computationally expensive to be viable. This computational cost is reduced using a combination of configuration interaction (CI) and perturbation theory (PT) which was first introduced in \cite{DBHF2017} and used in \cite{LDFDb2018, LDF118} for calculating the spectra of SHE Db ($Z=105$) and Og ($Z=118$). In this work we give a brief outline of the CIPT method and its implementation for the elements we calculate. For an in-depth discussion please refer to \cite{DBHF2017}. A fast version of this method has beed developed in~\cite{FCI}.

To generate the single-electron wavefunctions for all the elements,  we use the $V^{N_e-1}$ approximation (where $N_e$ is the total number of electrons) \cite{Kelly1964, Dzuba2005} where the Hartree-Fock calculations are performed for the singly-charged open-shell atom with a   $6d^n7s$ configuration, where $n=4,5,6$ and $7$ for Sg, Bh, Hs and Mt respectively. The  single-electron basis states are calculated in the field of the frozen atomic core. The basis sets are generated using a  B-spline technique~\cite{Johnson1988}  with 40 B-spline states in each partial wave of order 9 in a box with radius $40 \ a_B$ (where $a_B$ is the Bohr radius) with partial waves up to $l_{max}=4$ (where $l$ is orbital angular momentum) and the many-electron basis states $|i \rangle = \Phi_i(r_1,\dots,r_{N_e})$ (where $r_j$ is the radial position of the $j$th electron)  for the CI calculations are formed by making all possible single and double excitations from reference low-lying non-relativistic configurations of the atom. This set of many-body electron wavefunctions is ordered from lowest to highest energy and divided into two sets, 
\begin{itemize}
\item $P$: A small set of low energy wavefunctions ($i \leq N_{\text{Eff}}$, where $N_{\text{Eff}}$ is the number of wavefunctions in the low energy set) that give dominant contributions to the CI wavefunction.
\item $Q$: A large set of high energy wavefunctions ($N_{\text{Eff}}<i \leq N_{\text{total}}$) that are corrections to the wavefunctions from $P$.
\end{itemize}
The CI wavefunction is written as an expansion over single-determinant many-electron states $|i \rangle $ from these two sets,
\begin{equation}
| \Psi \rangle = 
\sum_{i=1}^{N_{\text{Eff}}} c_{i}|i\rangle + \sum_{i = N_{\text{Eff}} + 1}^{N_{\text{total}}} c_{i}|i\rangle .
\label{eq:psi}
\end{equation}
where $c_i$ are coefficients of expansion. The CI Hamiltonian is truncated by neglecting the off-diagonal matrix elements of the CI Hamiltonian between terms in $Q$ ($\langle i | H^{\text{CI}} | j \rangle = 0 $ for $|i\rangle, |j\rangle \in Q$),  which reduces the problem of finding the wave function and corresponding energy to a matrix eigenvalue problem of the size $P$ with modified CI matrix
\begin{equation} \label{eq:CI}
(H^{\rm CI} - EI)X=0,
\end{equation}
where $I$ is unit matrix, the vector $X = \{c_1, \dots, c_{N_{\rm eff}}\}$ and the low energy matrix elements of $H^{\rm CI}$ are modified to include perturbative contributions between states in $P$ and $Q$.
\begin{equation}
\langle i|H^{\rm CI}|j\rangle \rightarrow \langle i|H^{\rm CI}|j\rangle + 
\sum_k \frac{\langle i|H^{\rm CI}|k\rangle\langle k|H^{\rm
    CI}|j\rangle}{E - E_k}. 
    \label{eq:HCI}
\end{equation}
where $|i\rangle, |j\rangle \in P$, $|k\rangle \in Q$,  $E_k = \langle k|H^{\rm CI}|k\rangle$, and $E$ is the energy of the state of interest. 

Both the Breit interaction (magnetic interaction and retardation)\cite{Breit1929, Mann1971}  and quantum electrodynamic (QED) radiative corrections  (Ueling potential and electric and magnetic form factors) \cite{FG2005} are included in the calculations as described in our earlier works (see, e.g. \cite{FF113-115}).  As both the Breit and QED radiative corrections scale with atomic charge, $Z$ faster than the first power \cite{FF113-115}, their contribution to the energy levels of SHE is non-negligible. It was shown in \cite{LDFDb2018} that the magnitude of the combined correction to the energy levels of Db is at most  200~cm$^{-1}$.  A similar correction is expected for the SHE in this work.

For each level we calculate the Land\'{e} $g$-factor and compare it to the non-relativistic expression,
\begin{align} \label{eq:Lande}
g_{\text{NR}} =  1 + \dfrac{J(J + 1) - L(L+1) + S(S+1)}{2J(J+1)}.
\end{align}
Where possible, for each level we use $g_{\text{NR}}$ to find an analogous state in the lighter element to obtain an approximate label in the $LS$ coupling scheme. In fact, $LS$ notations do not make sense for the highly relativistic SHE states due to very large spin-orbit interaction (so the eigenvectors will look strongly mixed in $LS$ notation), we only use $LS$ notations for comparison with lighter elements. Otherwise, we label the $n$th sequential state of total angular momentum $J$ and parity by $n_{J}^{\text{parity}}$.

\section{Estimation of the accuracy} \label{sec:Accuracy}

Theoretical uncertainty is dominated by incomplete treatment of inter-electron correlations. These correlations can be further separated into core-valence and valence-valence correlations. We will discuss each of these separately. For the SHE calculations, the core includes all states in closed shells
from $1s$ to $5f$ containing one hundred electrons occupying one hundred states.
All other states, including states of the $6d$ and $7s$ shells are treated as
valence states. Only valence states are used in calculation of the CI matrix.
This means that we neglect core-valence correlations. To estimate the corresponding uncertainties, we perform calculations of the energy levels of gold and roentgenium (Rg, $Z$=111). Both these elements have one external electron above a closed $5d$ or $6d$ shell. We perform the calculations using the correlation potential method~\cite{Dzuba1988, DFSS1987_2}. In this method, core-valence correlation corrections are obtained using the electron self-energy operator (correlation potential) $\Sigma$\footnote{Do not confuse this with  the QED self-energy operator which we included using the radiative potential method \cite{FG2005}. This is the many-body self-energy operator, which for example, has been defined in the textbook \cite{LandauStatPhysPart2}. We calculate this operator using a Feynman diagram technique with relativistic Hartree-Fock Green's functions \cite{Dzuba1988}.} calculated by summation of the diagrams in the many-body perturbation theory.  The operator $\Sigma $ is defined by the correlation correction to the energy of the valence electron on the orbital $n$, $\delta E_n = \langle n | \Sigma | n \rangle$.   For the Au and Rg calculations, the upper complete $d$-shell ($5d$ or $6d$) is attributed to the core, and the correlation interaction of the external electron with the core is described by a correlation potential $\Sigma$.   \\

Calculation of $\Sigma$ involves a summation over all core states from
$1s$ to $5d$ for Au or $6d$ for Rg. This summation is strongly dominated by the upper $d$-shell. E.g., the $5d$ shell gives about 90\% of the correlation correction to the energies of the $6s$ and $6p$ valence states of Au, and more than 80\% of the correlation correction to the energies of the $7s$ and $7p$ valence states of Rg (see Table \ref{tab:CPaccuracy}). This is because of the small energy interval between the energies of
the $5d$ (or $6d$) state and the energies of lowest valence states.  Since correlation correction to the energy of the $s$ and $p$ valence states is about 20\%, the effect of neglecting inner-core contributions to the core-valence correlations is about 1 to 2\% of the energy of valence states.


\begin{table}
\center
\caption{Removal energies (cm$^{-1}$) for states of external electron of
Au and Rg calculated in different approximations. RHF is relativistic
Hartree-Fock, $\Sigma$($nd$) are Brueckner orbital energies calculated with correlation
potential $\Sigma$, in which summation over core states is limited to $5d$ or
$6d$ shell only. $\Sigma$(all) are the energies calculated with full summation over
core states. \label{tab:CPaccuracy}}
\begin{tabular}{l@{\hspace{0.7cm}}r@{\hspace{0.7cm}}r@{\hspace{0.7cm}}r@{\hspace{0.7cm}}r}
\toprule
\toprule
\multicolumn{5}{c}{Au} \\
  &  \centering    RHF &  $\Sigma$($5d$) &    $\Sigma$(all) &   Expt~\cite{NIST_ASD} \\
  \midrule
$6s_{1/2}$ &  60 179 &  75 539  & 77 878 &  74 409 \\
$6p_{1/2}$ &  29 303 & 36 508  & 37 322 & 37 051 \\
$6p_{3/2}$ &  26 664  & 32 314  & 32 785 & 33 324 \\
$6d_{3/2}$  & 11 929  & 12 423  & 12 439 & 12 457 \\
$6d_{5/2}$ &   11 875  & 12 344 &  12 357 & 12 376 \\
\\
\multicolumn{5}{c}{Rg} \\
    &   \centering  RHF  &    $\Sigma$($6d$) &   $\Sigma$(all)  \\
    \midrule
$7s_{1/2}$  & 83 436 & 101 901 & 106 780 \\
$7p_{1/2}$ &  38 006 &  49 996 &  52 269 \\
$7p_{3/2}$ &  26 550 &  33 659 &  34 685 \\
$7d_{3/2}$  & 11 859 &  12 594 &  12 656 \\
$7d_{5/2}$  & 11 738  & 12 383  & 12 428 \\
\bottomrule
\bottomrule
\end{tabular}
\end{table}

   It is interesting to note that in the second order of many-body perturbation theory, the correlation potential$\Sigma$ always overestimates the value of the correlation correction. This is because it does not include the effect of screening of inter-electron interaction by other atomic electrons. This effect appears in higher orders of perturbation theory. Its proper inclusion leads to very accurate results (see, e.g.~\cite{Dzuba1988, Dzuba2008}).
   
Since in the present work we do not go beyond the second order, we have the fortunate situation where neglecting inner-core contributions to $\Sigma$ has a similar affect on its value as the screening would do. In other words, the effect of neglecting the higher-order perturbative contributions on
the calculated energies partially compensates the effect of neglecting
screening of interelectron interactions. The data in Table \ref{tab:CPaccuracy} show that this is the case at least for the
$6s$ state of Au (and probably for the $7s$ state of Rg) where the correlation correction
is the largest in value. Therefore, it is reasonable to assume that the theoretical uncertainty is dominated by valence-valence correlations. The main source for it is the perturbative treatment of the excited configurations. The best way of estimating the uncertainty is to compare the theoretical and experimental energies for lighter elements. We did this in detail for W~\cite{DBHF2017}, which is the lighter analog of Sg, and for pairs Ta and Db~\cite{LDFDb2018}, and Rn and Og~\cite{LDFOg2018}. As follows from this comparison and from the analysis in Sections \ref{sec:Sg}, \ref{sec:Bh}, the theoretical uncertainty for the energies is on the level of $\sim$ 1000~cm$^{-1}$, sometimes a little higher (e.g.  $\sim$ 2000~cm$^{-1}$ for odd states of Bh). The uncertainty for ionization potentials is on the level of a few percent (see section \ref{sec:SHEIP}).

\section{Electric dipole transitions and isotope shifts} \label{sec:Isoshift}

In the spectroscopic measurements, the frequencies of strong electric dipole (E1) optical transitions ($\omega < 40 000$~cm$^{-1}$) are likely to be measured first as it has been done for the $^1$S$_0$ $\rightarrow$ $^1$P$_1^{\rm o}$ transition in No ($Z=102$)~\cite{Laatiaoui2016}. Broad spectrum scans for strong lines are unfeasible and therefore \textit{a priori} estimates of both a transition frequency and its strength from theoretical calculations will aid the experiments on SHE. Calculation of frequencies will be considered in section~\ref{sec:spectra}. In this work we also calculate the E1 transition amplitudes and rates for the major optical  transitions between the ground state and the  lowest states of opposite parity (odd states) for each of the four SHE of interest. 
\linebreak
To calculate the E1 transition amplitude $D_{\text{E1}}$ between two states $|a\rangle$ and $|b\rangle$, we use a self-consistent random-phase approximation (RPA) to simulate the atom in an external electromagnetic field. This results in an effective dipole field for the electrons that includes direct and exchange core polarization. An in-depth discussion of this method can be found in Ref.~\cite{DFSS1986, Dzuba2018}. The results in the  RPA approximation are gauge-invariant \cite{DFSS1986}. However, when you calculate correlation corrections beyond RPA, the length form of the E1 operator usually gives better results for low-frequency transitions. Indeed, the calculation of the correlation corrections can be made explicitly gauge-invariant in the case of one electron above closed shells \cite{DFSS1987_2, DFSS1987}. However, in the velocity form some correlation corrections are proportional to $1/\omega$ and become very large for small frequencies $\omega$  \cite{DFSS1987_2, DFSS1987}. This is the reason why we prefer to perform all calculations using the length form of the E1 operator. \\

Note that comparison of results in different gauges is not always a good test of accuracy. For example, in the RPA approximation and in the correlation potential approach described in Ref. \cite{DFSS1987_2, DFSS1987}, velocity and length forms give exactly the same results though the error is still finite. Therefore, to estimate the accuracy of the calculations we use comparison with available experimental data (see Table \ref{tab:E1_comp}).\\

The E1 transition rates, $A_{\text{E1}}$, are calculated using (in atomic units),
\begin{align} \label{eq:E1}
A_{\text{E1}} = \dfrac{4}{3}\left(\alpha \omega\right)^3\dfrac{ D_{\text{E1}}^2}{2J + 1}
\end{align}
where $J$ is the angular momentum of the upper state, $\alpha$ is the fine structure constant and $\omega$ is the frequency of the transitions in atomic units. All calculated amplitudes, $D_{\text{E1}}$, obey the selection rules for E1 transitions. The accuracy of these calculations cannot be tested directly due to the lack of experimental data on SHE and therefore we must rely on comparisons in lighter elements. Using the above method we calculated the E1 transition amplitudes and transition rates for the lighter analogs and compared them to available experimental data in Table \ref{tab:E1_comp}. The accuracy for the E1 amplitudes is $\sim$ 50\% which is sufficient to identify the strongest transitions. The calculated rates are \textit{ab initio} using the amplitudes and energies calculated in the CIPT method.  \\
\begin{table*}[h]
\center
\caption{Comparison of E1 transition amplitudes and rates between experimental and CIPT values for the lighter analogs of SHE, W \textsc{I}, Re \textsc{I}, Os \textsc{I}, and Ir \textsc{I}. Here $D_{\text{E1}}$, $A_{\text{E1}}$ and $gf$ are the transition amplitude, rate and oscillator strength respectively. The experimental E1 amplitudes were calculated using the experimental energies, transition rates from experimental sources and Eq.~(\ref{eq:E1}). To calculate oscillator strengths for comparison with Re I transitions from Ref. \cite{Ortiz2012},  we use the formula $gf = 3.062\times 10^{-6} \omega D^2_{\text{E1}}$ where $\omega$ is in cm$^{-1}$ and $D_{\text{E1}}$ is in (a.u.).  \label{tab:E1_comp}}
\begin{tabular}{c@{\hspace{0.5cm}}c@{\hspace{1cm}}c@{\hspace{0.5cm}}c@{\hspace{0.5cm}}c@{\hspace{0.5cm}}c@{\hspace{0.5cm}}c@{\hspace{0.5cm}}c@{\hspace{0.5cm}}}
\toprule
\toprule
 & \multicolumn{3}{c}{Expt.} & & \multicolumn{3}{c}{CIPT}\\
 \cline{2-4} \cline{6-8}
 \\
State & $E^{a}$  & $D_{\text{E1}}$ & $A_{\text{E1}}$ & & $E$& $D_{\text{E1}}$ & $A_{\text{E1}}$   \\
&  (cm$^{-1}$) & (a.u.) &  ($\times 10^{6} \ \text{s}^{-1}$) & &  (cm$^{-1}$) & (a.u.) &  ($\times 10^{6} \ \text{s}^{-1}$) \\
\hline
\multicolumn{8}{c}{W I} \\
13$_{1}^{\rm_o}$ & 39 183.19 & 2.09(9) &  178(15)$^{b}$ & & 39 606 &  3.07 & 400   \\
\\
\multicolumn{8}{c}{Os I} \\
4$_{4}^{\rm_o}$ & 32 684.61  & 2.00(7) & 31.53(221)$^{c}$ & & 32 576 &  2.36  & 43 \\
3$_{4}^{\rm_o}$ & 30 591.45 & 0.96 & 5.8$^{d}$  & & 30 359 &  1.37 & 12 \\
2$_{5}^{\rm_o}$ & 30 279.95 & 1.40(5) & 10.05(70) $^{c}$  & & 31 904 &  2.15 & 28 \\
\\
\multicolumn{8}{c}{Ir I} \\
3$_{11/2}^{\rm_o}$ & 39 940.37 & 1.72(22) & 32(8)$^{e}$ & & 41 083 & 1.49 & 26 \\
$^4$F$_{9/2}^{\rm_o}$ & 37 871.69 & 2.07(26) & 47(12)$^{e}$ & & 39 227 &  1.52 & 28 \\
$^4$D$_{7/2}^{\rm_o}$ & 37 515.32 & 1.73(22) & 40(10)$^{e}$ & & 40 106 &  1.55 & 39  \\
$^6$G$_{9/2}^{\rm_o}$ & 35 080.70 & 1.59(20) & 22(6)$^{e}$ & & 36 703 &  2.72 & 74 \\
$^6$G$_{11/2}^{\rm_o}$ & 34 180.46 & 1.45(7) & 14.2(14)$^{e}$ & & 36 358 & 1.51 & 18 \\
\midrule
State & $E^{a}$  & $D_{\text{E1}}$ & $gf$ & & $E$& $D_{\text{E1}}$ & $gf$   \\
&  (cm$^{-1}$) & (a.u.) &   & &  (cm$^{-1}$) & (a.u.) &  \\
\midrule
 \multicolumn{8}{c}{Re I} \\
 $^6$P$_{3/2}^{\rm_o}$ & 28 961.55 & 1.22(17) & 0.132(36)$^{f}$ & & 29 303 &  1.80 & 0.29    \\
 $^6$P$_{7/2}^{\rm_o}$ & 28 889.72 &  2.26(27) & 0.45(11)$^{f}$ & & 29 247 &  3.32 &  0.98  \\
  $^6$P$_{5/2}^{\rm_o}$ & 28 854.18 &  1.70(19)  & 0.254(56)$^{f}$ & & 29 505 & 2.51 & 0.57   \\
\bottomrule
\bottomrule
\end{tabular}
\begin{flushleft}
$^a$ Ref. \cite{NIST_ASD}, $^b$ Ref. \cite{Kling1999}, $^c$ Ref. \cite{Ivarsson2003},    $^d$ Ref. \cite{Kwiatkowski1984},  $^e$ Ref. \cite{Fuhr1996}, $^{f}$ Ref. \cite{Ortiz2012}
\end{flushleft}
\end{table*}
Along with the excitation spectrum and E1 transitions we also calculate the isotope shift (IS) for each transition. \\
\linebreak
The IS is the difference in the transition frequency between two different isotopes. The IS is important for at least two reasons. First, it can be used to find the difference in nuclear radius between two isotopes. Second, it can be used to predict the spectra of heavier, meta-stable neutron rich isotopes from the spectra of short-lived, neutron deficient isotopes created and measured in the laboratory. These predictions can be compared to astronomical data \cite{DFW17, Polukhina2012, Gopka2008, Fivet2007} and could lead to the discovery of isotopes in the ``island of stability'' where it is expected that meta-stable, neutron-rich isotopes are created in cosmological events\cite{Goriely2011, Fuller2017, Friebel2018, Schuetrumpf2015}. The  IS of SHE is strongly dominated by the volume shift (also known as ``field shift'' in literature~\cite{Stacey1966}), while the mass shift is negligible. Using CIPT, we calculate the excitation spectrum of the each isotope by varying the nuclear radius in the HF procedure described in the previous section.  In the zero approximation only $s_{1/2}$ and $p_{1/2}$ electron waves penetrate the nucleus and for these the dependence of IS on the nuclear radius $R_N$ is  $R_N^{2 \gamma}$ where  $\gamma = \sqrt{1-(Z\alpha)^2}$ - see details in Ref.  \cite{FGV2018}. Higher waves undergo isotopic shifts due to change of the $s_{1/2}$ and $p_{1/2}$  wave functions and corresponding changes in the atomic Hartree-Fock potential - the core relaxation effect. Therefore, the dependence of the field IS on the nuclear radius in any atomic transition in multi-electron atoms is always $R_N^{2 \gamma}$. Using the large-scale trend for nuclear radii $R_N \propto A^{1/3}$  the isotopic volume shift can be also approximated by $\delta \nu \propto A^{2\gamma/3}$ \cite{DFW17, FGV2018}  as nuclear shell fluctuations are suppressed \cite{Angeli2013}.  The first form of the IS we present is given by
\begin{align} \label{eq:isoa}
\delta \nu &= E_{2} - E_{1} = a\left(A_{2}^{2\gamma/3} - A_{1}^{2\gamma/3}\right),
\end{align}
where $A_1$ and $A_2$ are atomic numbers for two isotopes ($A_2>A_1$), $E_1$ and $E_2$ are the excitation energy for  $A_1$ and $A_2$ respectively and $a$ is a parameter which should be calculated for each transition. This form of the IS is convenient for non-neighbouring isotopes and predicting the spectra of meta-stable isotopes because there is a significant difference in the values of $A$ for isotopes synthesized in laboratory and hypothetical meta-stable isotopes ($\Delta A \sim 10$). The $R_N \propto  A^{1/3}$  trend is based on the constant nuclear density approximation due to finite range nuclear interactions. Variation of the nuclear shape and charge density may lead to significant deviations. Specific theoretical information about expected density distributions in SHE is presented in \cite{Nazarewicz2018}. 

 A more common form of isotope shift is the standard formula relating the change of atomic frequency to the change of nuclear charge radius
\begin{align} \label{eq:isoF}
\delta \nu &= F\delta \left<r^{2}\right>,
\end{align}
where the square of the nuclear charge radius is calculated using the Fermi distribution for the nuclear density. This formula (neglecting the mass shift) is convenient for extraction of the nuclear charge radius change from isotope shift measurements of nearby isotopes. Lastly, we introduce a new form of the IS which should be valid for all isotopes. Using  the RMS (root mean squared) nuclear radius, $ R_{rms} = \sqrt{\left<r^{2}\right>}$,  and $\delta \nu \propto \delta R_{rms}^{2\gamma}$ \cite{FGV2018} we can write the equation,
\begin{align}\label{eq:isoFtilde}
\delta \nu = \tilde{F}\dfrac{R_{rms,A_2}^{2\gamma} - R_{rms,A_1}^{2\gamma}}{\text{fm}^{2\gamma}}
\end{align}
where $\tilde{F}$ is an IS parameter to be calculated for each transition.
\section{Calculation of energy levels, E1 transition rates and isotope shift} \label{sec:spectra}

\begin{figure*}
\center
\includegraphics[scale=1]{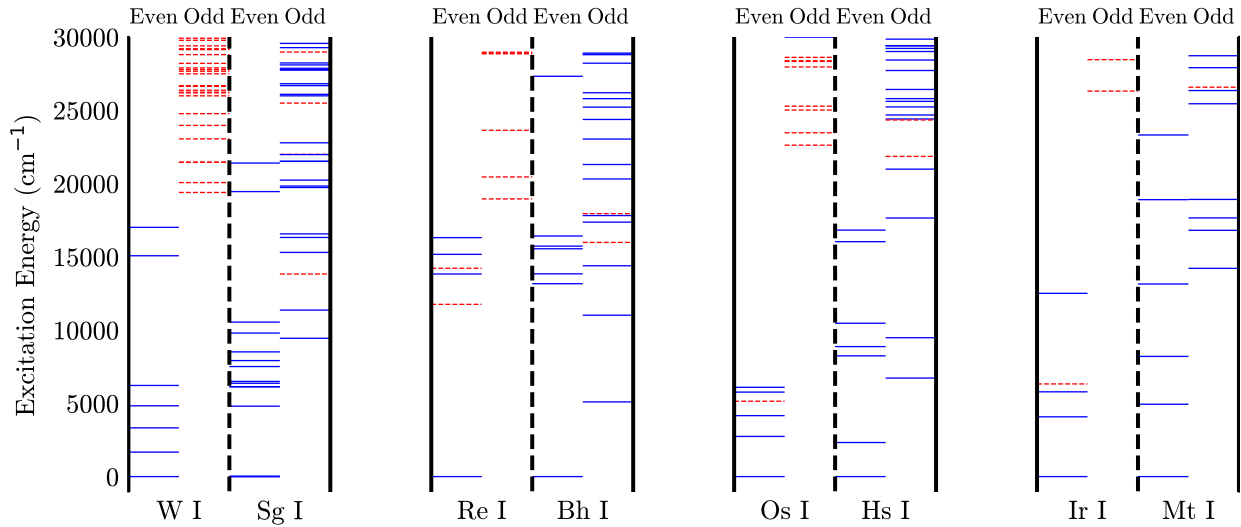} 
\caption{Comparison of low-energy excitations of SHE and their respective lighter analogs. For each element, the states are split between odd and even parities. The solid (blue) lines represent states with $6s^2$ or $7s^2$ in the electronic configuration for the lighter elements and SHE respectively. The dashed (red) lines are all other states where an $s$ electron has been excited from the filled $6s$ or $7s$ shell. Experimental energies were used for W \textsc{i}, Bh \textsc{i}, Hs \textsc{i} and Ir \textsc{i}. \cite{NIST_ASD}\label{fig:EL}}
\end{figure*}

Energy levels of SHE are calculated by solving the matrix eigenvalue problem (\ref{eq:CI}) separately for states of given value of the total angular momentum $J$ and parity. The specific details for each considered SHE are presented below. Most previous theoretical works on these SHE present the calculation of the first ionization potential, which we discuss in Section~\ref{sec:SHEIP}. Fig.~\ref{fig:EL} compares calculated spectra of low-lying states of SHE with experimental data on their lighter analogs. One can see a significant difference in the spectra of SHE and their lighter analogs, which is common for all considered atoms. Almost all low-lying odd states of lighter atoms correspond to the $6s-6p$ excitation from the ground state. In contrast to that, in SHE the $7s$ state is significantly lower on the energy scale than the $6d$ state due to relativistic effects. Therefore, dominant excitations occur from the $6d$ state, i.e. low-lying odd states correspond to the $6d-7p$ excitations from the ground state. Since the $6d-7p$ energy interval is smaller than the $6s-6p$ one, the density of odd states is higher for the SHE.

\subsection{The Seaborgium Atom} \label{sec:Sg}

\begin{table*}[t] 
\caption{Low-energy spectrum of even- and odd-parity states for Sg \textsc{i}, Bh \textsc{i}.   We present the energy and Land\'e $g$-factor for each state $J^{\text{parity}}$. We present $LS$- notations only for comparison with lighter analogs.  For SHE states where an analogous state cannot be found in the lighter analog the term is labeled according to the sequential number of the state ($n$) for the given $J^{\text{parity}}$ group, $n_{J}^{\text{parity}}$.\label{tab:SHESpectrumSgBh}}
 		\center 
 		\begin{tabular}{cl@{\hspace{0.5cm}}c@{\hspace{0.5cm}}r@{\hspace{0.5cm}}r@{\hspace{1cm}}l@{\hspace{0.5cm}}c@{\hspace{0.5cm}}r@{\hspace{0.5cm}}r} 
 		\toprule 
 \toprule 
&  \multicolumn{4}{c}{Sg \textsc{i}} &  		\multicolumn{4}{c}{Bh \textsc{i}} \\
 \cmidrule{2-5} \cmidrule{6-9} 
& \parbox{2cm}{Major \\ Configuration} & Term  &   \parbox{1cm}{Energy \\ (cm$^{-1}$)}  &  \parbox{1.2cm}{Land\'{e} \\$g$-factor}  & \parbox{2cm}{Major \\ Configuration} & Term &   \parbox{1cm}{Energy \\ (cm$^{-1}$)}  &  \parbox{1.2cm}{Land\'{e} \\$g$-factor}  \\ 
 		\midrule 
\multicolumn{9}{c}{Even parity states}\\
 (1) &  $6d^4 7s^2$ &  $^5$D$_0$    & 0 & 0.00  &  $6d^5 7s^2$  &  $^6$S$_{5/2}$    & 0 & 1.78 \\ 
 (2) &  $6d^4 7s^2$ &  $^5$D$_1$     & 4 834 & 1.50  &   $6d^5 7s^2$  &  $^4$P$_{3/2}$    & 13 062 & 1.32 \\ 
 (3) &  $6d^4 7s^2$ &  $^5$D$_2$    & 7 614 & 1.44  &  $6d^5 7s^2$  &  $^4$G$_{7/2}$    &  13 828 & 1.15 \\ 
 (4) &  $6d^4 7s^2$ &  $^5$D$_3$   & 9 607 & 1.39  &   $6d^5 7s^2$  &  $^4$G$_{11/2}$    &  14 981 & 1.19 \\  
 (5) &   $6d^4 7s^2$ &  $^5$D$_4$    & 10 335 & 1.27 &   $6d^5 7s^2$  &  $^4$P$_{1/2}$   &  15 659 & 1.90 \\ 
 (6) &  $6d^4 7s^2$ &  $^5$2P$_0$    & 13 592 & 0.00  &   $6d^5 7s^2$  &  $^4$G$_{9/2}$    & 16 447 & 1.17 \\  
\multicolumn{9}{c}{Odd parity states}\\
(7) &  $6d^3 7s^2 7p$  & 1$_2^{\rm_o}$    & 14 717 & 0.57  &   $6d^4 7s^2 7p $  &  $^2$S$_{1/2}^{\rm_o}$    & 12 792 & 0.72 \\  
(8) &   $6d^3 7s^2 7p$  &  1$_1^{\rm_o}$    & 17 043 & 0.71 &   $6d^4 7s^2 7p $   &  $^6$D$_{1/2}^{\rm_o}$   & 17 781 & 1.66 \\  
 (9) &  $6d^3 7s^2 7p$  & 2$_2^{\rm_o}$    & 20 444 & 1.13 &   $6d^4 7s^2 7p $   &  $^6$D$_{3/2}^{\rm_o}$    & 19 483 & 1.09 \\  
(10) &   $6d^3 7s^2 7p$  & 1$_3^{\rm_o}$    & 20 628 & 0.97  &   $6d^5 7s 7p$  &  $^8$P$_{5/2}^{\rm_o}$    & 22 228 & 2.08 \\  
(11) &  $6d^4 7s 7p$  &  $^7$F$_0^{\rm_o}$   & 20 979 & 0.00 &   $6d^4 7s^2 7p $   &  $^6$P$_{3/2}^{\rm_o}$    & 22 533 & 1.74 \\  
(12) &  $6d^3 7s^2 7p$  & 2$_1^{\rm_o}$     & 22 041 & 2.02  &   $6d^4 7s^2 7p $   &  $^6$D$_{5/2}^{\rm_o}$    & 22 930 & 1.26 \\  
 (13) &  $6d^3 7s^2 7p$  &  1$_4^{\rm_o}$   & 24 132 & 1.11 &   $6d^5 7s 7p$  &  $^8$P$_{7/2}^{\rm_o}$    & 24 020 & 1.67 \\ 
(14) &  $6d^3 7s^2 7p$  & 3$_1^{\rm_o}$    & 24 382 & 1.19  &   $6d^4 7s^2 7p $   &  $^6$D$_{7/2}^{\rm_o}$    & 25 171 & 1.28 \\  
(15) &  $6d^4 7s 7p$  &  $^1$S$_0^{\rm_o}$   & 25 362 & 0.00 &   $6d^4 7s^2 7p $   &  $^6$D$_{9/2}^{\rm_o}$    & 26 587 & 1.21 \\  
 (16) &  $6d^3 7s^2 7p$  & 2$_3^{\rm_o}$  & 25 966 & 1.29  &   $6d^4 7s^2 7p $  &  $^6$F$_{5/2}^{\rm_o}$    & 28 060 & 1.57 \\  
 (17) &  $6d^3 7s^2 7p$  &  1$_{5}^{\rm_o}$   & 26 271 & 1.17  &  $6d^4 7s^2 7p $  & 3$_{1/2}^{\rm_o}$  & 29 823 & 0.44 \\ 
(18) &   $6d^3 7s^2 7p$  & 3$_2^{\rm_o}$     & 26 420 & 1.22 &   $6d^4 7s^2 7p $ &  3$_{3/2}^{\rm_o}$  & 29 885 & 1.55 \\
(19) &  $6d^4 7s 7p$  &  $^7$F$_1^{\rm_o}$  & 27 030 & 1.40  &  $6d^4 7s^2 7p $  & 3$_{7/2}^{\rm_o}$  & 31 078 & 1.24 \\ 
(20) & $6d^3 7s^2 7p$  & 4$_2^{\rm_o}$   & 27 416 & 1.74  &   $6d^4 7s^2 7p $ & 4$_{5/2}^{\rm_o}$  & 31 253 & 1.30 \\ 
(21) & $6d^4 7s  7p $  &  $^7$F$_2^{\rm_o}$   & 29 976 & 1.41   &   $6d^4 7s^2 7p $  & 4$_{7/2}^{\rm_o}$  & 32 814 & 1.37 \\ 
(22) & $6d^3 7s^2 7p$  &  3$_0^{\rm_o}$ & 30 055 & 0.00 &  $6d^4 7s^2 7p $  & 4$_{3/2}^{\rm_o}$  & 33 459 & 1.40 \\
(23) & $6d^3 7s^2 7p$  &  2$_4^{\rm_o}$ & 30 372 & 1.25 & $6d^4 7s^2 7p $  & 2$_{9/2}^{\rm_o}$  & 33 575 & 1.06 \\
(24) & $6d^3 7s^2 7p$ & 3$_3^{\rm_o}$  & 30 753 & 1.09 & $6d^4 7s^2 7p $  & 5$_{5/2}^{\rm_o}$     & 33 738 & 1.04 \\
(25) & $6d^3 7s^2 7p$  & 5$_1^{\rm_o}$  & 30 868 & 0.92 & $6d^4 7s^2 7p $  &  4$_{1/2}^{\rm_o}$  & 35 408 & 2.22 \\
(26) & $6d^3 7s^2 7p$ & 4$_3^{\rm_o}$ & 31 647 & 1.34  & $6d^4 7s^2 7p $  & 5$_{3/2}^{\rm_o}$   & 35 447 & 1.00 \\
(27) & $6d^3 7s^2 7p$  & 6$_2^{\rm_o}$  & 32 040 & 1.13 & $6d^4 7s^2 7p $  & 6$_{5/2}^{\rm_o}$  & 35 774 & 1.34 \\
(28) & $6d^3 7s^2 7p$  &  3$_4^{\rm_o}$ & 32 073 & 1.00 & $6d^4 7s^2 7p $  & 5$_{7/2}^{\rm_o}$   & 36 251 & 1.00 \\
(29) & $6d^3 7s^2 7p$ &  4$_0^{\rm_o}$ & 32 381 & 0.00 & $6d^4 7s^2 7p $  & 6$_{3/2}^{\rm_o}$    & 36 333 & 1.02 \\
(30) & $6d^3 7s^2 7p$ &  6$_1^{\rm_o}$  & 32 520 & 1.21 & $6d^4 7s^2 7p $  & 7$_{5/2}^{\rm_o}$    & 36 875 & 1.25 \\
(31) & $6d^4 7s  7p $  &  $^5$D$_3^{\rm_o}$ & 32 885 & 1.47 & $6d^4 7s^2 7p $  &  1$_{11/2}^{\rm_o}$    & 37 542 & 1.10 \\
(32) & $6d^3 7s^2 7p$  &  4$_4^{\rm_o}$ & 33 339 & 1.23 & $6d^4 7s^2 7p $  & 6$_{7/2}^{\rm_o}$  & 37 910 & 1.32 \\
(33) & $6d^3 7s^2 7p$  & 7$_2^{\rm_o}$  & 33 602 & 1.08 & $6d^4 7s^2 7p $  & 7$_{3/2}^{\rm_o}$     & 37 954 & 1.05 \\
(34) & $6d^3 7s^2 7p$  & 8$_2^{\rm_o}$  & 34 147 & 1.45 & $6d^5 7s 7p $  &  $^8$P$_{9/2}^{\rm_o}$    & 37 972 & 1.62 \\
(35) & $6d^3 7s^2 7p$  &  2$_{5}^{\rm_o}$ & 34 380 & 1.12 & $6d^4 7s^2 7p $  & 4$_{9/2}^{\rm_o}$  & 38 336 & 1.23 \\  
(36) & $6d^3 7s^2 7p$  & 6$_3^{\rm_o}$  & 34 538 & 1.13 & $6d^4 7s^2 7p $  & 8$_{5/2}^{\rm_o}$  & 39 454 & 1.19 \\
(37) & $6d^3 7s^2 7p$  & 7$_1^{\rm_o}$  & 35 110 & 1.42 & $6d^4 7s^2 7p $  & 7$_{7/2}^{\rm_o}$  & 39 602 & 1.33 \\
(38) & $6d^3 7s^2 7p$  & 7$_3^{\rm_o}$  & 35 897 & 1.31 & $6d^4 7s^2 7p $  & 5$_{1/2}^{\rm_o}$  & 40 273 & 1.76 \\
(39) & $6d^3 7s^2 7p$  &  5$_4^{\rm_o}$ & 36 629 & 1.29 & \\
(40) & $6d^3 7s^2 7p$  & 9$_2^{\rm_o}$  & 36 695 & 1.24 \\
(41) & $6d^3 7s^2 7p$ & 8$_3^{\rm_o}$  & 36 846 & 1.18 \\
(42) & $6d^3 7s^2 7p$  & 8$_1^{\rm_o}$  & 37 169 & 1.30 \\
(43) & $6d^3 7s^2 7p$  &  6$_4^{\rm_o}$& 37 218 & 1.25 \\
(44) & $6d^3 7s^2 7p$  &  3$_{5}^{\rm_o}$ & 37 542 & 1.26 \\
(45) & $6d^3 7s^2 7p$  &  5$_0^{\rm_o}$  & 38 322 & 0.00 \\ 
(46) & $6d^3 7s^2 7p$  & 9$_3^{\rm_o}$  & 38 547 & 1.12 \\ 
(47) & $6d^3 7s^2 7p$  & 10$_2^{\rm_o}$  & 38 915 & 1.22 \\ 
(48) & $6d^3 7s^2 7p$  &  7$_4^{\rm_o}$ & 39 138 & 1.30 \\
(49) & $6d^3 7s^2 7p$  &  4$_{5}^{\rm_o}$ & 39 337 & 1.23 \\
(50) & $6d^3 7s^2 7p$  & 10$_3^{\rm_o}$  & 39 725 & 1.23 \\
(51) & $6d^3 7s^2 7p$  & 9$_1^{\rm_o}$  & 40 073 & 1.62 \\ 
  \bottomrule
 \bottomrule
 \end{tabular} 
 \end{table*} 
Seaborgium was first experimentally detected in 1974 \cite{Ghoirso1974}. Since the initial discovery there has been continued interest and study into its physical and chemical properties including the discovery of isotopes with longer lifetimes. There exist some experimental results for Sg \textsc{i} in the field of chemistry \cite{Schadel2012}. However, there are no spectroscopic results available.   The ground state configuration of Sg \textsc{i} is expected to be [Rn]$5f^{14}6d^47s^2$, similar to the ground state of its lighter homologue (W \textsc{i}, ground configuration:  [Xe]$4f^{14}5d^46s^2$). \\
\linebreak
We calculated the first 6 even parity states and the ground state was found  to be the [Rn]$5f^{14}6d^47s^2 \ ^5$D$_0$ state. To calculate the even states we use three reference configurations, $6d^47s^2$, $6d^5 7s$ and $6d^6$ to make states in the effective CI matrix (first terms in the expansion (\ref{eq:psi}) and in the CI effective Hamiltonian (\ref{eq:HCI})). All other states, which are treated as corrections to the states from reference configurations (second terms in the expansion (\ref{eq:psi}) and in the CI Hamiltonian (\ref{eq:HCI})) are obtained by exciting one or two electrons from the reference configurations.
Similarly, for odd parity states we use the reference states from the $6d^47s7p$, $6d^37s^27p$ and $6d^57p$ configurations.  All calculated even and odd energy levels are presented in Table \ref{tab:SHESpectrumSgBh}. Similar calculations were performed for W~\textsc{i}  using analogous reference states and the same parameters. Comparing these results to the experimental spectrum~\cite{NIST_ASD} we found a maximum discrepancy of $|\Delta| \approx 600 $~cm$^{-1}$ and expect a similar accuracy for our Sg~\textsc{i} calculations. Note that this accuracy is slightly better than what was reported in Ref.~\cite{DBHF2017} due to inclusion of a larger number of states into the effective CI matrix.

Comparing the spectrum of Sg~\textsc{i} in Table~\ref{tab:SHESpectrumSgBh} to the spectrum of W~\textsc{I}~\cite{NIST_ASD}, we can see the manifestation of relativistic effects. As discussed above, relativistic effects cause the $7s$ orbital in Sg \textsc{i} to be strongly contracted and more tightly bound in comparison to the $6s$ orbital in W~\textsc{I}. The same effects also push out the $6d$ orbital of Sg \textsc{i} in comparison to the $5d$ orbital in W \textsc{i}. In the  W~\textsc{i} spectrum there are low-lying states corresponding to the $6s \rightarrow 5d$ excitation from the ground state (e.g., the $5d^56s \ ^7$S$_3$ state at 2 951.29 cm$^{-1}$). In contrast, in the Sg~\textsc{i} spectrum, all low-lying even states belong to the  $6d^4 7s^2$ configuration. The relativistic effects are more apparent in the low-lying odd parity states of Sg~\textsc{i}. In W~\textsc{i} all odd states correspond to the $6s \rightarrow 6p$ excitation from the ground state, while in Sg~\textsc{i} most of the low-lying odd states correspond to the $6d \rightarrow 7p$ excitation.
Only a few of the Sg \textsc{i} predicted in the optical region correspond to the $7s \rightarrow 7p$ excitation. 

We calculate rates of electric dipole transitions from the ground state to excited states of the opposite parity using the approach described in Section~\ref{sec:Isoshift}. The results are presented in Table~\ref{tab:SHEE1transitionSgBh}. There are not many such transitions due to the zero value of the total angular momentum $J$ in the ground state. Because of that, the transitions are only allowed to the odd states with $J=1$. A few transitions are good candidates for the detection. The transition with the highest transition rate is $^5$D$_0 \rightarrow 9_1^{\rm_o}$ ($\omega = 40 \ 073 \text{ cm}^{-1}$).

We also present the isotopic shift parameters, $F$ and $a$ from equations (\ref{eq:isoa}) and (\ref{eq:isoF}), in Table~\ref{tab:SHEE1transitionSgBh} for each respective E1 transition. The two isotopes we use are $^{269}$Sg  and $^{290}$Sg ($R_{rms,\text{269}} = 5.8814$ fm and $R_{rms,\text{290}}  = 6.0145$ fm respectively),  where $^{290}$Sg is the theoretically  metastable ($N=184$) isotope of Sg. 

\subsection{The Bohrium Atom}  \label{sec:Bh}

 Bohrium was first discovered in 1981 \cite{Munzenberg1981}. No atomic spectra have been measured or calculated for any Bh isotopes or ions. When calculating the energy spectrum of Bh \textsc{i}, we use a similar approach as with Sg \textsc{i}.  For the low-lying even parity spectrum we use an effective CI matrix build from the states of the $6d^5 7s^2$, $6d^6 7s$ and $6d^7$ reference configurations. For the odd parity spectrum we use the states from the $6d^5 7s 7p$, $6d^4 7s^2 7p$ and $7d^6 7p$ reference configurations. The lowest six even parity states and low-lying odd parity states are presented in Table~\ref{tab:SHESpectrumSgBh}. For an estimate of accuracy we calculated the low-lying spectrum of Re~\textsc{i} (the lighter analogue of Bh) with similar parameters. Comparing the CIPT calculated spectrum to the experimental spectrum \cite{NIST_ASD}, the energy discrepancy (with respect to the ground state) was $\Delta \approx 900$~cm$^{-1}$ for the even parity states, while for the odd parity states $\Delta \approx 2000$~cm$^{-1}$.\\
\linebreak
 The calculated Bh~\textsc{i} ground state is $6d^5 7s^2 \ ^6$S$_{5/2}$.  As with Sg~\textsc{i},  we see the relativistic effect of the tightly bound $7s$ electron which results in the primary excitation of the $6d$ electron. Comparing the spectrum of Bh~\textsc{i} with that of Re~\textsc{I} in Fig. \ref{fig:EL} we see that there are several low-lying states in Re~\textsc{I} corresponding to $6s \rightarrow 5d$ excitations (the lowest is at $11 754.52$~cm$^{-1}$), while there are no similar low-lying states in Bh~\textsc{i}.  The density of low-energy odd-parity states is much larger in Bh~\textsc{i} than in Re~\textsc{i}. The Bh \textsc{i} low odd-parity states are completely dominated by the $6d \rightarrow 7p$ excitations in calculated spectrum and and there are no  $7s \rightarrow 7p$ excitations. 
 The odd-parity state comparison between Bh~\textsc{i} and Re~\textsc{I} is similar to that of Sg~\textsc{i} and W~\textsc{i} in Section \ref{sec:Sg}. In the spectrum of Re~\textsc{i}~\cite{NIST_ASD} there do exist states corresponding to  $5d \rightarrow 6p$ transitions from the ground state; however, they occur much higher in the spectrum compared to Bh~\textsc{i} where the $6d \rightarrow 7p$ excitations dominate. It should be noted that the number of low-lying odd-parity states is larger in Bh~\textsc{i} than in Re \textsc{i}. The lowest odd state of Bh \textsc{i} occurs at $12 792$~cm$^{-1}$, whereas in Re~\textsc{i} the lowest odd state is at $18 950$~cm$^{-1}$.\\
 \linebreak
 Bh \textsc{i} has a large number of allowed low-energy optical E1 transitions from the ground state, which are presented  in Table \ref{tab:SHEE1transitionSgBh}. The isotope shift parameters, $a$ and $F$, are calculated using formulas (\ref{eq:isoa}) and (\ref{eq:isoF}) after calculating the atomic spectra for the theoretically meta-stable isotope of $^{270}$Bh using the CIPT method. We use the values of RMS nuclear radii  $ R_{rms,\text{270}} =  5.8879$ fm for $^{291}$Bh and $R_{rms,\text{291}} = 6.0207$ fm  for $^{291}$Bh.
 
\subsection{The Hassium atom} \label{sec:Hs}

Hassnium ($Z=108$) was first synthesized in 1984~\cite{Munzenberg1984}. We present the low-lying levels and the first ionization energy of Hs~\textsc{i} in Table~\ref{tab:SHESpectrumHsMt}. For the low-lying even spectrum effective CI reference states belong to the $6d^5 7s^2$, $6d^6 7s$ and $6d^7$ configurations. For the odd spectrum we use reference states of the $6d^5 7s 7p$, $6d^4 7s^2 7p$ and $7d^6 7p$ configurations.  Note that the half-filled $6d$ sub-shell makes computational methods particularly expensive. However, using the CIPT method the computation becomes tractable.  

Once again it is interesting to compare the spectra of Hs~\textsc{i} with the analogue Os~\textsc{i} in the period above. In the even states of Os~\textsc{i} there are states corresponding to the $6s \rightarrow 5d$ excitations from the ground state. In the Hs~\textsc{i} spectrum all low-lying even states belong to the  $6d^5 7s^2$ configuration. No states with the $7s \rightarrow 6d$ excitation were found. The odd states are similar to those of Sg and Bh, with the primary excitation $6d \rightarrow 7p$ in Hs~\textsc{i} while there are no  $5d \rightarrow 6s$ excitations in low Os~\textsc{i} spectrum. The odd states of Hs~\textsc{i} also lie much lower than those in Os~\textsc{i}.  The lowest odd state of Hs~\textsc{i} is 13 949 cm$^{-1}$, while the first odd state of Os~\textsc{i} occurs at 22 615.69 cm$^{-1}$ \cite{NIST_ASD}. 

The allowed strong optical E1 transitions from the low-lying odd states to the ground state ($^5$D$_{4}$) are presented in Table~\ref{tab:SHEE1transitionSgBh}. As with Bh \textsc{i} there is a large number of strong optical transitions. The transition with the largest rate is 3$_5^{\rm o}$ $\rightarrow$ $^5$D$_{4}$ ($\omega =$~39~268~cm$^{-1}$). Other possibly detectable transitions include 5$_3^{\circ}$ $\rightarrow$ $^5$D$_{4}$ ($\omega =$~34~812~cm$^{-1}$) and 2$_5^{\circ}$ $\rightarrow$ $^5$D$_{4}$ ($\omega= $~34~739~cm$^{-1}$).

We also present the isotopic shift parameters for the Hs E1 optical transitions in Table~\ref{tab:SHEE1transitionSgBh}. These were calculated from the theoretical spectra (calculated with the CIPT method) with isotopes $^{270}$Hs and $^{292}$Hs with RMS nuclear radii $R_{rms,\text{270}} = 5.8879$ fm for $^{292}$Hs and $R_{rms,\text{292}} = 6.0207$ fm for $^{270}$Hs.

\subsection{The Meitnerium Atom} \label{sec:Mt}

		\begin{table*}[t] 
\caption{Low-lying spectrum of even and odd states parity for Hs \textsc{i} and Mt \textsc{i}.   We present the energy and Land\'e g-factor for each state $J^{\text{parity}}$. We present $LS$- notations only for comparison with lighter analogs. For SHE states where an analogous state cannot be found in the lighter analog the term is labeled according to the sequential number of the state ($n$) for the given $J^{\text{parity}}$ group, $n_{J}^{\text{parity}}$.\label{tab:SHESpectrumHsMt}}
 		\center 
 		\begin{tabular}{cl@{\hspace{0.5cm}}c@{\hspace{0.5cm}}r@{\hspace{0.5cm}}r@{\hspace{1cm}}l@{\hspace{0.5cm}}c@{\hspace{0.5cm}}r@{\hspace{0.5cm}}r} 
 		\toprule 
 \toprule  		
& \multicolumn{4}{c}{Hs \textsc{i}} &  		\multicolumn{4}{c}{Mt \textsc{i}} \\
 \cmidrule{2-5} \cmidrule{6-9} \\
& \parbox{2cm}{Major \\ Configuration} & Term &   \parbox{1cm}{Energy \\ (cm$^{-1}$)}  &  \parbox{1.2cm}{Land\'{e} \\g-factor}  & \parbox{2cm}{Major \\ Configuration} & Term &   \parbox{1cm}{Energy \\ (cm$^{-1}$)}  &  \parbox{1.2cm}{Land\'{e} \\g-factor}  \\ 
 		\midrule 
  		 	\multicolumn{8}{c}{Even parity states}\\
 (1) &  $6d^6 7s^2$  &  $^5$D$_{4}$   & 0 & 1.37  & $6d^7 7s^2$  &  $^4$F$_{9/2}$ & 0 & 1.265 \\ 
 (2) & $6d^6 7s^2$  &  $^5$D$_2$   & 2 102 & 1.38  &  $6d^7 7s^2$  &  $^4$F$_{3/2}$  & 5 047 & 1.214 \\  
 (3) &  $6d^6 7s^2$  &  $^5$D$_{0}$ & 7 400 & 0.00  &   $6d^7 7s^2$  &  $^4$F$_{5/2}$ & 7 996 & 1.222 \\ 
 (4) &  $6d^6 7s^2$  &  $^5$D$_{3}$ & 8 270 & 1.43  &   $6d^7 7s^2$  &  $^4$F$_{7/2}$ & 12 628 & 1.213 \\ 
 (5) &   $6d^6 7s^2$  &  $^5$D$_1$   & 9 285 & 1.41  &   $6d^7 7s^2$  &  $^2$G$_{3/2}$ & 17 368 & 0.931 \\ 
 (6) &    $6d^6 7s^2$  &  $^3$H$_{5}$ & 15 816 & 1.11 & $6d^7 7s^2$  &  $^2$G$_{5/2}$   & 18 467 & 1.409\\ 
\multicolumn{8}{c}{Odd parity states}\\
  (7) &   $6d^5 7s^2 7p$  & 1$_2^{\rm_o}$    & 13 093 & 1.98 &   $6d^6 7s^2 7p $  & 1$_{7/2}^{\rm_o}$   & 21 879 & 1.44 \\  
  (8) &  $6d^5 7s^2 7p$  & 1$_3^{\rm_o}$     & 15 600 & 1.58  &   $6d^6 7s^2 7p $  &  1$_{9/2}^{\rm_o}$ & 24 388 & 1.33 \\ 
 (9) &  $6d^5 7s^2 7p$  & 2$_2^{\rm_o}$  & 23 708 & 1.30 &   $6d^6 7s^2 7p $  & 1$_{3/2}^{\rm_o}$  & 24 524 & 1.51 \\ 
 (10) &  $6d^5 7s^2 7p$  & 2$_3^{\rm_o}$     & 26 492 & 1.16  &   $6d^6 7s^2 7p $ & 1$_{5/2}^{\rm_o}$   & 25 990 & 1.25 \\ 
(11) &   $6d^6 7s 7p$  &  $^7$D$_{4}^{\rm_o}$   & 27 394 & 1.58 &   $6d^6 7s^2 7p $  & 2$_{5/2}^{\rm_o}$   & 31 975 & 1.54 \\  
(12) &   $6d^5 7s^2 7p$  &  1$_1^{\rm_o}$   & 29 444 & 1.17 &  $6d^6 7s^2 7p $  &1$_{1/2}^{\rm_o}$    & 32 851 & 0.81 \\  
(13) &    $6d^5 7s^2 7p$  &  3$_2^{\rm_o}$    & 29 794 & 1.34 &  $6d^7 7s  7p$  &  $^{6}$D$_{9/2}^{\rm_o}$ & 33 505 & 1.40 \\ 
(14) &   $6d^6 7s 7p$  &  $^7$D$_5^{\rm_o}$    &30 863 & 1.37   &  $6d^6 7s^2 7p $  & 2$_{1/2}^{\rm_o}$   & 34 665 & 1.51\\  
(15) &  $6d^5 7s^2 7p$  & 3$_3^{\rm_o}$     & 30 908 & 1.32  &   $6d^6 7s^2 7p$  & 2$_{7/2}^{\rm_o}$    & 35 117 & 1.29 \\ 
(16) &   $6d^5 7s^2 7p$  & 4$_2^{\rm_o}$    & 31 165 & 1.33 &   $6d^6 7s^2 7p$  & 2$_{3/2}^{\rm_o}$     & 36 159 & 1.13 \\ 
(17) &   $6d^5 7s^2 7p$  & 2$_4^{\rm_o}$      & 31 295 & 1.40  &   $6d^7 7s 7p$  &  $^6$F$_{11/2}^{\rm_o}$   & 38 027 & 1.31 \\  
(18) &  $6d^5 7s^2 7p$  & 1$_0^{\rm_o}$  & 31 552 & 0.00 &   $6d^6 7s^2 7p$& 3$_{7/2}^{\rm_o}$   & 38 450 & 1.17 \\ 
(19) &  $6d^5 7s^2 7p$ & 3$_4^{\rm_o}$     & 32 522 & 1.26 & $6d^6 7s^2 7p$  & 3$_{9/2}^{\rm_o}$    & 39 296 & 1.13 \\ 
(20) &  $6d^5 7s^2 7p$ & 5$_2^{\rm_o}$   & 33 694 & 1.44 & $6d^6 7s^2 7p$  & 2$_{11/2}^{\rm_o}$     & 41 310 & 1.33 \\ 
(21) &  $6d^5 7s^2 7p$ & 4$_3^{\rm_o}$     & 33 920 & 1.03 \\ 
(22) &  $6d^5 7s^2 7p$ & 2$_1^{\rm_o}$      & 34 076 & 1.52 \\ 
(23) &  $6d^5 7s^2 7p$  & 2$_5^{\rm_o}$     & 34 739 & 1.20 \\ 
(24) &  $6d^5 7s^2 7p$ & 5$_3^{\rm_o}$   & 34 812 & 1.41 \\
(25) &  $6d^5 7s^2 7p$ & 4$_4^{\rm_o}$   & 35 689 & 1.23 \\ 
(26) &  $6d^6 7s 7p$  &  $^7$D$_{3}^{\rm_o}$  & 35 705 & 1.56 \\ 
(27) &  $6d^5 7s^2 7p$  & 3$_1^{\rm_o}$   & 35 990 & 1.81 \\ 
(28) &  $6d^6 7s 7p$  &  $^7$D$_2^{\rm_o}$  & 37 036 & 1.40 \\ 
(29) &  $6d^6 7s 7p$  & $^7$P$_{3}^{\rm_o}$ & 37 237 & 1.33 \\ 
(30) &  $6d^5 7s^2 7p$ & 5$_4^{\rm_o}$   & 37 443 & 1.18 \\ 
(31) &  $6d^5 7s^2 7p$ & 7$_2^{\rm_o}$   & 38 519 & 1.34 \\
(32) &  $6d^6 7s 7p$  &  $^7$P$_{4}^{\rm_o}$ & 39 025 & 1.29 \\ 
(33) &  $6d^5 7s^2 7p$  & 3$_5^{\rm_o}$   & 39 268 & 1.27 \\
(34) &  $6d^6 7s 7p$  &  $^7$D$_1^{\rm_o}$  & 39 512 & 2.11 \\ 
(35) &  $6d^5 7s^2 7p$  & 8$_3^{\rm_o}$  & 39 652 & 1.38 \\ 
(36) &  $6d^5 7s^2 7p $  & 9$_3^{\rm_o}$    &  40 783  &  1.19 \\
  \bottomrule
 \bottomrule
 \end{tabular} 
 \end{table*} 
Meitnerium ($Z=109$) was first synthesized in 1982 \cite{Munzenberg1982}. The ground state of Mt~\textsc{i} is expected to follow that of the element in the above period  (Ir) with [Rn]$5f^{14}6d^{7}7s^2 \ ^4$F$_{9/2}$ which we confirm in the calculated spectrum presented in Table \ref{tab:SHESpectrumHsMt}.\\

We use the same method as for previous elements to calculate the low-lying spectrum of Mt~\textsc{i}. We present the lowest six even states using the  $6d^7 7s^2$, $6d^8, 7s$ and $6d^9$ reference configurations. We also present the first 12 odd parity states for which the $6d^7 7s 7p$, $6d^6 7s^2 7p$ and $6d^8 7p$ configurations were used. The results are in Table~\ref{tab:SHESpectrumHsMt}. Comparison with lighter analog Ir~\textsc{i} shows similar trend as for other SHE Sg, Bh and Hs. 

 We also present the allowed E1 transitions for Mt and the respective isotope shift parameters in Table~ \ref{tab:SHEE1transitionSgBh}. The high energy of the odd states in Mt~\textsc{i} result in a small number of allowed E1 transitions within optical region from the ground state compared to Bh and Hs.  Promising transitions for future measurement include $^6$F$_{11/2}^{\circ} \rightarrow ^4$F$_{9/2}$ ($\omega =$~38~027~cm$^{-1}$) and $^{6}$D$_{9/2}^{\circ} \rightarrow ^4$F$_{9/2}$ ($\omega =$ 33~505~cm$^{-1}$). All other rates are two or more orders of magnitude smaller. For the synthesized and metastable isotopes we use the RMS nuclear radii values, $R_{rms,\text{276}} = 5.9265$ fm  and $R_{rms,\text{293}} = 6.0330$ fm.
 
  \begin{table*}[t] 
 \caption{Strong electric dipole transitions and isotopic shift parameters for Sg \textsc{i}, Bh \textsc{i}, Hs~\textsc{i} and Mt~\textsc{i}. Only direct optical transitions to the ground state satisfying the E1 transition selection rules are shown. Here $D_{\text{E1}}$ is the transition amplitude in a.u., $A_{\text{E1}}$ is the transition rate, $a$, $F$, and $\tilde{F}$ are calculated isotopic shift parameters for the charge radius discussed in Section \ref{sec:Isoshift}. The numbers in parentheses correspond to the numbered states in Tables \ref{tab:SHESpectrumSgBh} and \ref{tab:SHESpectrumHsMt}  for the respective element. \label{tab:SHEE1transitionSgBh}}
\begin{tabular}{l@{\hspace{0.01cm}}c@{\hspace{0.5cm}}r@{\hspace{0.5cm}}r@{\hspace{0.5cm}}r@{\hspace{0.5cm}}r@{\hspace{0.5cm}}r|l@{\hspace{0.05cm}}cr@{\hspace{0.5cm}}r@{\hspace{0.5cm}}r@{\hspace{0.5cm}}r@{\hspace{0.5cm}}r@{\hspace{0.25cm}}r}  
\toprule
\toprule
& State &   \parbox{1cm}{$D_{\text{E1}}$ \\ (a.u)} & \parbox{1cm}{$A_{\text{E1}}$ \\ { \small $(\times 10^{6} \ \text{s}^{-1})$ }} & \parbox{1cm}{$a  $ \\ (cm$^{-1}$)} & \parbox{1cm}{$F $ \\ ($\frac{\text{cm}^{-1}}{\text{fm}^{2}}$)} &   \multicolumn{1}{c}{\parbox{1cm}{$\tilde{F} $ \\ (cm$^{-1}$) }} & & State &   \parbox{1cm}{$D_{\text{E1}}$ \\ (a.u)} & \parbox{1cm}{$A_{\text{E1}}$ \\ $(\times 10^{6} \ \text{s}^{-1})$ } & \parbox{1cm}{$a  $ \\ (cm$^{-1}$)} & \parbox{1cm}{$F $ \\ ($\frac{\text{cm}^{-1}}{\text{fm}^{2}}$)} & \multicolumn{1}{c}{\parbox{1cm}{$\tilde{F} $ \\ (cm$^{-1}$) }}  \\
\midrule
 		\multicolumn{7}{c}{Sg \textsc{i} (Ground state: $^5$D$_0$)} &  \multicolumn{7}{c}{Bh \textsc{i} (Ground State: $^6$S$_{5/2}$)} \\
 		\\
(8) &  1$_1^{\rm_o}$        & 0.639 & 1.36 & 9.41& 2.04 & 11.9  &(9)  & $^6$D$_{3/2}^{\rm_o}$     & -0.172 & 0.107 & 18.1 & 3.74 & 22.8  \\ 
(12) & 2$_{1}^{\rm_o}$          & -0.160 & 0.192 & -2.95 & -0.639 & -3.73 & (10) & $^8$P$_{5/2}^{\rm_o}$     & -0.474 &  0.812 & 83.4 & 17.2 & 105 \\ 
(14) & 3$_{1}^{\rm_o}$         & 1.17  & 13.4 & 4.90 & 1.06 & 6.18 & (11) &  $^6$P$_{3/2}^{\rm_o}$     & -0.494 & 1.38 & -101 & -20.7 & -127 \\ 
(19) & $^3$P$_1^{\rm_o}$      & -0.163 & 0.353 & -19.7 & -4.25 & -24.8 & (12) &  $^6$D$_{5/2}^{\rm_o}$     & -0.0391 & 0.00611 & -120 & -24.6 & -151 \\ 
(25) & 5$_{1}^{\rm_o}$   & 0.592 & 6.97 & 6.58 & 1.42 & 8.30 & (13)  & $^8$P$_{7/2}^{\rm_o}$     & 0.500 & 0.858 & 84.5 & 17.4 & 107 \\ 
(30) &  6$_{1}^{\rm_o}$     & -0.412 & 3.95 & 7.01 & 1.52 & 8.85  & (14)  & $^6$D$_{7/2}^{\rm_o}$     & 0.345 & 0.471 & -63.3 & -13.0 & -79.7 \\ 
(37) &  7$_{1}^{\rm_o}$ & -0.302 & 2.67&  1.66 & 0.36 & 2.10 & (16) & $^6$F$_{5/2}^{\rm_o}$     & 1.51 & 16.6 & -160 & -33.0 & -202 \\
 (42) & 8$_{1}^{\rm_o}$   & 0.148& 0.761 & 3.55 & 0.768 & 4.48 & (18)  & 3$_{3/2}^{\rm_o}$  & 1.50 & 30.0 & -64.9 & -13.4 & -81.7 \\
(51) & 9$_{1}^{\rm_o}$   & 0.524 & 11.9 & -4.77& -1.03& -6.01  & (19) & 3$_{7/2}^{\rm_o}$   & 1.75 & 23.3 & 44.0 & 9.06& 55.4 \\
\multicolumn{6}{c}{} & &  (20) & 4$_{5/2}^{\rm_o}$   &  -0.433 & 1.90 & -101 & -20.7 & -127  \\
\multicolumn{6}{c}{} & &  (21) & 4$_{7/2}^{\rm_o}$   &  1.88 & 31.2 & -380 & -78.4 & -479 \\
\multicolumn{6}{c}{} & & (22) & 4$_{3/2}^{\rm_o}$  &   -0.998 & 18.6 & -41.3 & -8.51& -52 \\
\multicolumn{6}{c}{} & & (24) & 5$_{5/2}^{\rm_o}$  &   -0.101 & 0.131 & -105 & -21.6 & -132 \\
\multicolumn{6}{c}{} & & (26) & 5$_{3/2}^{\rm_o}$   &   0.438 & 4.27 & -135 & -27.9 & -170 \\
\multicolumn{6}{c}{} & & (27) & 6$_{5/2}^{\rm_o}$  &   -1.06 & 17.1 & -364 & -74.9 & -458 \\
\multicolumn{6}{c}{} & & (28) & 5$_{7/2}^{\rm_o}$  &  0.0665 & 0.0361 & -34.7 & -7.15 & -43.7 \\
\multicolumn{6}{c}{} & & (29) & 6$_{3/2}^{\rm_o}$    &  0.160 & 0.615 & -70.6 & -14.5 & -88.9 \\
\multicolumn{6}{c}{} &  & (30) & 7$_{5/2}^{\rm_o}$    &  -0.539 & 4.86 & -335 & -69.0 & -422\\
\multicolumn{6}{c}{} & & (33) & 6$_{7/2}^{\rm_o}$   &   -0.674 & 6.18 & -129 & -26.6 & -163 \\
\multicolumn{6}{c}{} & &  (34) & 7$_{3/2}^{\rm_o}$   &  0.387 & 4.09 & -513 & -106 & -647 \\
\multicolumn{6}{c}{} & &  (37) & 8$_{5/2}^{\rm_o}$   &  0.232 & 1.10 & -561 & -116 & -707\\
\multicolumn{6}{c}{} & &  (38) & 7$_{7/2}^{\rm_o}$  &  0.516 & 4.13 & -364 & -75.1 & -459 \\
\midrule
 		\multicolumn{7}{c}{Hs \textsc{i} (Ground State: $^5$D$_{4}$)} & \multicolumn{7}{c}{Mt \textsc{i} (Ground State: $^4$F$_{9/2}$)}\\
 		\\
(8)  & 1$_{3}^{\rm_o}$     & 0.501 &  0.276 & 22.7 & 4.45 & 28.9& (7)	& 1$_{7/2}^{\rm_o}$	  & 0.0537 & 0.00765 & 27.5& 5.10 & 34.5 \\
(10) & 2$_{3}^{\rm_o}$      & 0.224 & 0.269 & 22.9 & 4.49 & 28.8 &   (8)	& 1$_{9/2}^{\rm_o}$	 & 0.432 & 0.550 & 27.6 & 5.13 & 34.7 \\
(11) & $^7$D$_{4}^{\rm_o}$    & -1.11 & 5.66 & -29.1 & -5.70 & -36.6  & (13)	&	$^{6}$D$_{9/2}^{\rm_o}$  & 1.27 & 12.3 & -51.7 & -9.60 & -64.9  \\
(14) & $^7$D$_5^{\rm_o}$     & 0.999 &  5.41& -26.2 & -5.15 & -33.0 &  (15) & 2$_{7/2}^{\rm_o}$	   & -0.294 & 0.946 & 33.3 & 6.18 & 41.8 \\
(15) & 3$_{3}^{\rm_o}$     & 0.208 & 0.370 & 16.2 & 3.18 & 20.4  & (17)	&	$^6$F$_{11/2}^{\rm_o}$    & -1.89 & 33.3 & -47.9& -8.89& -60.1 \\
(17) & 2$_{4}^{\rm_o}$      & 0.0934 & 0.0603 & 5.54 & 1.09 & 6.98 &  (19) 	& 3$_{9/2}^{\rm_o}$	    & 0.0954 & 0.112 & 19.0 & 3.53 & 23.9  \\ 
(19) & 3$_{4}^{\rm_o}$      & 0.120 & 0.112 & 18.5 & 3.62 & 23.2 & (20) 	& 2$_{11/2}^{\rm_o}$	    & 0.170 & 0.344 & 25.7 & 4.78 & 32.3  \\
(21) & 4$_{3}^{\rm_o}$      & -0.150 & 0.253 & 20.7 & 4.05 & 26.0 & \\
(23) & 2$_{5}^{\rm_o}$      & -1.13 & 9.88 & 12.3 & 2.42 & 15.5   &\\
 (24) & 5$_{3}^{\rm_o}$    & 1.70 & 35.5 & 12.3 & 2.41 & 15.5  &\\
(25) & 4$_{4}^{\rm_o}$      & 0.798 & 6.52 & 7.84 & 1.54 & 9.87  &\\ 
(26)  &  $^7$D$_{3}^{\rm_o}$   & -0.493 & 3.20 & -33.3 & -6.53 & -41.9  &\\ 
(29) & $^7$P$_{3}^{\rm_o}$  & -0.511 & 3.91 & -15.8 & -3.11 & -19.9  &\\ 
(30) &5$_{4}^{\rm_o}$     & -0.297 & 1.04 & 1.95 & 0.382 & 2.45  & \\ 
(32) & $^7$P$_{4}^{\rm_o}$  & 0.425 &  2.41 & -9.56 & -1.87 & -12.0 &\\ 
(33) &3$_{5}^{\rm_o}$     & 2.64 & 77.5 & 5.16 & 1.01 & 6.49  & \\ 
(35) & 7$_{3}^{\rm_o}$    & 2.10 & 80.0 & -2.30 & -0.451 & -2.89  &\\ 
\bottomrule
\bottomrule
\end{tabular}

\end{table*}

\section{Ionization potentials and comparison with other data.} \label{sec:SHEIP}

As well as calculating the spectrum of neutral Sg, Bh, Hs and Mt we also calculated their first ionization potentials (IPs). To calculate the IP for each atom we use the same single-electron basis set for a neutral atom and an ion. The ionization potential is found as a difference between ground state energies of the atom and its ion. The effective CI matrix was built from all states of the $6d^n 7s$, $6d^{n-1}7s^2$ and $6d^{n+1}$ reference configurations ($n=4-7$ for Sg through to Mt). States that were treated perturbatively were obtained by exciting one or two electrons from the reference configurations and generating all single-determinant states from these configurations. We start from calculating the IPs of lighter analogs of the SHE to compare them with experiment. The results are in Table~\ref{tab:IP}. We also include in the table the results of the multi-configuration Dirac-Fock (MCDF) calculation~\cite{MCDF-Sg,MCDF-BhHs}. We do this because similar MCDF calculations have been used for the SHE (see Tables~\ref{tab:EE} and \ref{tab:SHEIP}). The CIPT values of the IPs agree with experiment within few percent (error $<$1\% for Ta, W, and Re, and $\sim$ 3\% for Os and Ir). We expect similar accuracy for the first IPs of SHE analogs presented in Table \ref{tab:SHEIP}.  For comparison, the difference between MCDF values of IPs of W, Re and Os and experimental IPs is larger than 10\% (Table~\ref{tab:IP}).


\begin{table}[h]
\caption{Theoretical and experimental ionization potentials of open $5d$-shell elements.  The CIPT energies are the results of the present work. \label{tab:IP}}
\begin{tabular}{llcccc}
\toprule
\toprule
           &                  & \multicolumn{3}{c}{IP (eV)} \\
Atom &  \parbox{1cm}{Ionic \\ State} & $J$ &  Expt.~\cite{NIST_ASD}   & CIPT  & MCDF \\
\midrule
Ta    & $5d^3 6s$  & 1   & 7.549    & 7.57  & \\
W     & $5d^4 6s$ & 1/2 & 7.864   & 7.90  &  6.97~\cite{MCDF-Sg}  \\
Re    & $5d^5 6s$ & 3   &  7.833  &  7.85  &  6.84~\cite{MCDF-BhHs} \\
Os    & $5d^6 6s$ & 9/2 & 8.438  &  8.69  &  7.45~\cite{MCDF-BhHs} \\
Ir      & $5d^7 6s$ & 5    & 8.967 &  9.27  & \\
\bottomrule
\bottomrule
\end{tabular}
\end{table}

Table~\ref{tab:EE} shows some resonance (corresponding to strong electric dipole transitions from the ground state) excitation energies for SHE and their lighter analogs calculated in the present work and by the MCDF method~\cite{MCDF-Sg,MCDF-BhHs}. The energies for lighter elements are compared to experiment. Our values are taken from Tables~\ref{tab:SHESpectrumSgBh} and \ref{tab:SHESpectrumHsMt}; for the MCDF energies we present all results which can be found in~\cite{MCDF-Sg,MCDF-BhHs}. There is  a significant difference in the excitation energies of SHE, while for lighter atoms the difference is not so large. There is a $\sim$ 10\% difference from experiment in both calculations. There are too little data on the MCDF calculations to come to any conclusion about the reasons for the differences. 
\begin{table}[h]
\caption{Some excitation energies (cm$^{-1}$) in open $6d$-shell SHE and their lighter analogs. The CIPT energies are the results of the present work.   \label{tab:EE}}
\begin{tabular}{lllccc}
\toprule
\toprule
Atom &  \multicolumn{2}{c}{State} &  Expt.~\cite{NIST_ASD}   & CIPT  & MCDF~\cite{MCDF-Sg,MCDF-BhHs} \\
\midrule

W  & $5d^46s^2$ & $^5$D$_1$ & 1670 &1502 & 1162 \\
      &                      & $^5$D$_2$ & 3325 & 2664 & 2581 \\

Re    & $5d^5 6s6p$ &  $^8$P$^{\rm o}_{5/2}$   &  18950  &    & 14000  \\

Os    & $5d^6 6s6p$ &  $^7$D$^{\rm o}_{5}$  & 23463  &  26000  &  20500 \\

Sg  & $6d^47s^2$ & $^5$D$_1$ &         &4834 & 4186 \\
      &                      & $^5$D$_2$ &        & 7614 & 7211 \\

Bh    & $6d^5 7s7p$ &  $^8$P$^{\rm o}_{5/2}$   &    & 2220   & 15100  \\

Hs    & $6d^5 7s^27p$ &  $^5$S$^{\rm o}_{2}$  &   &  13100  &  5100 \\
        &                           &  $^5$D$^{\rm o}_{3}$  &   &  15600  &  8600 \\

\bottomrule
\bottomrule
\end{tabular}
\end{table}

Finally, Table~\ref{tab:SHEIP} shows IPs of SHE and their ions. We included the result of our previous work on Db~\cite{LDFDb2018} together with the relativistic Hartree-Fock (RHF) calculations which include semi empirical core-polarisation correction~\cite{Dzuba2016} and the MCDF results~\cite{MCDF-Sg,MCDF-BhHs}. There are two sets of MCDF results. One, in the column marked as MCDF, is what directly comes from the MCDF calculations. We also presented prediction of MCDF IPs corrected by extrapolation of the difference with experiment from lighter atoms (marked as ``Extrap.''). As one can see from Table~\ref{tab:IP} the MCDF method tends to underestimate IPs by about 10\%. Therefore, multiplying the calculated IPs by a factor $\sim$ 1.1 extrapolated from lighter elements leads to better prediction of the IPs for SHE. Indeed, the extrapolated values are in better agreement with our CIPT calculations. Note however that the extrapolation assumes similarities between involved elements. In fact, they are significantly different. Ionization of lighter elements goes via removal of the $s$ electron ($6s$ electron for W, Re and Os). In contrast, ionization of SHE goes via removal of the $6d$ electron. RHF calculations (see Ref.~\cite{Dzuba2016} and Table~\ref{tab:SHEIP}) used a different type of extrapolation. Instead of extrapolating a final number, a term in the Hamiltonian was extrapolated. A term, simulating the effect of core polarisation, was added to the RHF Hamiltonian in Ref. \cite{Dzuba2016}. Its strength was chosen to fit IPs of lighter atoms. Then the same term was used for SHE. 

\begin{figure}
\center
\includegraphics[scale=1.00]{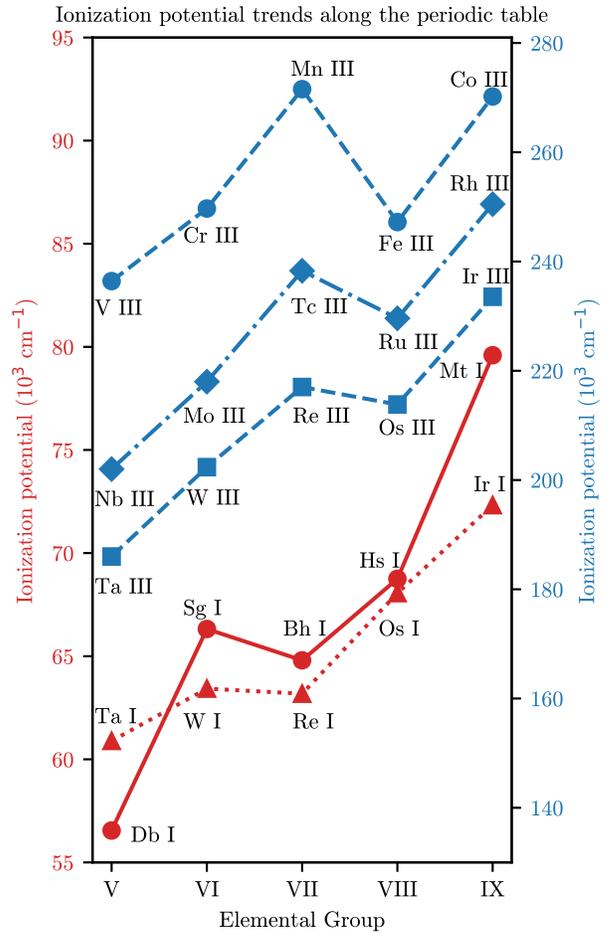}
\caption{Plot of ionization trends for open $d$-shell elements. The IP trend lines for the doubly ionized elements (blue) use the scale on the right and the neutral IP trend lines (red) use the left. The IPs of neutral SHEs and the doubly ionized lighter homologues Ta~\textsc{iii}, W~\textsc{iii}, Re~\textsc{iii}, Os~\textsc{iii} and Ir~\textsc{iii} were calculated using the CIPT method. All other IPs are from Ref. \cite{NIST_ASD}. \label{fig:IPPlot}}
\end{figure}

Studying IPs of SHE with open $6d$-shell shows a significant difference in trends compared to their lighter analogs. These differences are convenient to discuss by looking at the diagram in Fig.~\ref{fig:IPPlot}. The diagram shows trends in IPs of SHE with open $6d$-shell from Db to Mt together with the trends for lighter atoms from Ta to Ir. IPs for doubly ionized ions of lighter elements are also shown because they do not have external $s$-electrons, and further ionization of these ions goes via removal of a $d$-electron similar to what takes place for SHE. 

First, we note that the change of IPs from Ta to Ir is smooth and almost monotonic, apart from a small local minimum at Re atom. It shows increasing of IP towards the fully filled $5d$ shell. The ionization occurs via removal of a $6s$ electron. The $6s$ orbital is not very sensitive to the details of energy structure of other shells, which explains the smooth behaviour of the IP trend. In contrast, ionization of the SHE occurs via removal of a $6d$ electron. Strong relativistic effects manifest themselves in the trend of the IP change. A local maximum of the IP occurs for Sg atom that has four $6d$ electrons in the fully occupied $6d_{3/2}^4$ subshell. Removing an electron from a closed shell is difficult, therefore there is a local maximum. The next atom, Bh, has one more $6d$ electron, which has to occupy the $6d_{5/2}$ state. Due to large relativistic effects in SHE, there is a large fine structure interval between the  $6d_{3/2}$ and $6d_{5/2}$ states and therefore a significantly smaller IP for Bh (see Fig.~\ref{fig:IPPlot} and Table~\ref{tab:SHEIP}). A similar effect is known for an open $p$ shell where it is more pronounced. E.g., the IP of Bi, which has three $6p$ electrons, is smaller than for Pb, which has two $6p$ electrons corresponding to the closed $6p_{1/2}^2$ subshell. The effect is much more pronounced for SHE with an open $7p$ shell~\cite{FF113-115}. The IP of Mc ($Z$=115), which has three $7p$ electrons, is about 1.5 times smaller than the IP of Fl ($Z$=114), which has two $7p$ electrons.

To see whether a similar effect can be found in lighter atoms, we studied IPs of doubly ionized ions with an open $d$-shell (from $3d$ to $5d$). The ions were chosen because they do not have external $s$-electrons, and further ionization goes via removal of a $d$-electron. The results are shown in Fig.~\ref{fig:IPPlot}. Most IP values are taken from the NIST database~\cite{NIST_ASD}. However, NIST data for ions from Ta~III to Ir~III have poor accuracy. Therefore, we recalculated the IPs using the CIPT method. IPs of these ions show a different trend compared to the SHE. The maximum binding energy and hence the maximum IP is for a half filled $d$-shell in agreement with the non-relativistic Hund rule, which states that the maximum energy corresponds to the maximum possible value of the total spin.  This holds even for the heaviest of the three groups of ions. Thus, the SHE elements with the open $6d$ shell represents the only known example of a strong manifestation of relativistic effects, making  the energy difference between the $6d_{3/2}$ and $6d_{5/2}$ states more important than Hund's rule.

A similar manifestation of relativistic effects can be found in the trends of further ionization of the SHE ions (see Table~\ref{tab:SHEIP}). In many cases (e.g., the Bh and Hs ions) ionization from the $6d$ shell stops as soon as the fully filled $6d_{3/2}^4$ subshell is reached. Further ionization occurs from the $7s$ subshell.

\begin{table}[h]
\caption{Ionization potentials of open $6d$-shell SHE, including ions.  The CIPT energies are the results of the present work. \label{tab:SHEIP}}
\begin{tabular}{llccccc}
\toprule
\toprule
           &                  & \multicolumn{4}{c}{IP (eV)} \\
 \parbox{1cm}{Atom \\ or ion} &  \parbox{1cm}{Ground \\ State} & $J$ & CIPT  & RHF\footnotemark[1]  & \parbox{1cm}{MCDF \\ \cite{MCDF-Sg,MCDF-BhHs}} & \parbox{1cm}{Extrap. \\ \cite{MCDF-Sg,MCDF-BhHs}} \\
\midrule
Db~I   &  $6d^3 7s^2$ & 2    & 7.01 & 6.75 &  &  \\
&&&&&&\\
Sg~I    & $6d^4 7s^2$ & 0   & 8.22 &  7.70 & 7.03  & 7.85 \\
Sg~II   & $6d^3 7s^2$ & 3/2 & 18.0 &          & 15.85 &  17.06 \\
Sg~III  & $6d^2 7s^2$ &    2 & 24.8 &          & 24.61 &  25.74 \\
&&&&&&\\
Bh~I    & $6d^5 7s^2$ & 5/2 & 8.03 &  8.63  & 6.82  & 7.7 \\
Bh~II   & $6d^4 7s^2$ & 0    & 19.0 &           & 16.55  & 17.5 \\
Bh~III  & $6d^4 7s$    & 1/2 & 26.2 &            & 25.64  & 26.6 \\
Bh~IV  & $6d^4$        & 0    & 36.8 &            & 36.33  & 37.3 \\
&&&&&&\\
Hs~I    & $6d^6 7s^2$ & 4   & 8.52 &  9.52   & 6.69  & 7.6 \\
Hs~II   & $6d^5 7s^2$ & 5/2 & 19.7 &           & 16.62  & 18.2 \\
Hs~III  & $6d^4 7s^2$ & 3   & 27.7 &            & 27.12  & 29.3 \\
Hs~IV  & $6d^4 7s$    & 1/2 & 40.5 &           & 36.59  & 37.7 \\
Hs~V   & $6d^4 $        & 0    & 50.6 &           & 50.37  & 51.2 \\
&&&&&&\\
Mt~I    & $6d^7 7s^2$ & 9/2 & 9.86 & 10.4     &   & \\
Mt~II   & $6d^6 7s^2$ & 4 & 20.7 &             &   & \\
Mt~III  & $6d^5 7s^2$ & 5/2 & 28.4 &             &   & \\
Mt~IV  & $6d^5 7s$    & 3 & 43.3 &             &   & \\
Mt~V    & $6d^5$        & 5/2 & 50.3 &             &   & \\
\bottomrule
\bottomrule
\footnotetext[1]{Relativistic Hartree-Fock with semi-empirical core polarisation correction~\cite{Dzuba2016}}
\end{tabular}
\end{table}

\section{Conclusion}

Calculation of atomic spectra and optical E1 transitions for the elements in the superheavy region with open $d$-shells is novel. In spite of the extreme computational cost of existing methods, using perturbation theory we can calculate the low-lying energy states and relevant E1 transitions with a modest computational cost and with a small loss in accuracy \cite{DBHF2017}. In this work we presented the low-lying energy states for Sg \textsc{i}, Bh \textsc{i}, Hs \textsc{i} and Mt \textsc{i} including the optical transitions between the ground state and states of the opposite parity and their ionization potentials. For all SHEs we observed the relativistic effects, which contract the spectrum compared to their lighter analogs. This is advantageous as it results in a large number of states in the allowed E1 optical region and therefore enhances the likelihood of future measurements. These calculations will help to facilitate future experimental measurements of atomic spectra of these elements. We also presented the relevant isotopic field shift for optical E1 transitions for all four considered SHE. This may help the interpretation of future measurements and contribute to our understanding of the nuclear properties of elements in the superheavy region and potentially identify the existence of meta-stable superheavy isotopes in astronomical spectra.\\

\bibliographystyle{apsrev4-1}
\bibliography{SHE_106_109,other,super}

\end{document}